\documentclass[aps,nofootinbib,amsmath,prc,showpacs,groupedaddress,12pt,superscriptaddress,notitlepage]{revtex4-1}

\usepackage[paperwidth=210mm,paperheight=297mm,centering,hmargin=2.3cm,tmargin=2.8cm,bmargin=3.4cm]{geometry}

\usepackage{amsfonts}
\usepackage{amsmath}
\usepackage{epsfig}
\usepackage{latexsym}
\usepackage{paralist}
\usepackage{fancyhdr}
\usepackage{graphicx}

\usepackage{subfigure}                    
\usepackage{pgfplots}

\usepackage{soul}

\numberwithin{equation}{section}

\usepackage[vcentermath]{youngtab}
\usepackage{young}
\usepackage{ytableau}
\usepackage{etex}
\usepackage{braket}
\usepackage{float}
\usepackage{slashed}
\usepackage{xcolor}

\usepackage{dcolumn}

\makeatletter
\@addtoreset{equation}{section}

\makeatother

\def\bea{\begin{eqnarray}} 
\def\eea{\end{eqnarray}}
\def\be{\begin{equation}} 
\def\ee{\end{equation}} 
\def\ba{\begin{array}}
\def\ea{\end{array}}

\def\be{\begin{equation}}
\def\ee{\end{equation}}
\def\bea{\begin{eqnarray}}
\def\eea{\end{eqnarray}}

\renewcommand{\thefootnote}{\fnsymbol{footnote}}

\let\oldtitle\title
\renewcommand{\title}[1]{\oldtitle{\color{blue}{#1}}}

\let\oldeqref\eqref
\let\oldcite\cite
\renewcommand{\eqref}[1]{{\color{blue}\oldeqref{#1}}}
\renewcommand{\cite}[1]{{\color{blue}\oldcite{#1}}}

\makeatletter
\let\reftagform@=\tagform@
\def\tagform@#1{\maketag@@@{\ignorespaces\textcolor{blue}{(\ignorespaces #1 \unskip\@@italiccorr \ignorespaces)\ignorespaces}}}
\makeatother

\makeatletter
\renewcommand{\p@subsection}{}
\renewcommand{\p@subsubsection}{}
\makeatother

\usepackage{mathpazo}
\usepackage{amsmath}

\begin{document}


\title{Scaling solutions in the derivative expansion}

\author{N. Defenu}
\email{defenu@thphys.uni-heidelberg.de}
\affiliation{Institut f\"ur Theoretische Physik, Universit\"at 
Heidelberg, D-69120 Heidelberg, Germany}

\author{A. Codello}
\email{codello@cp3-origins.net}
\affiliation{CP$^3$-Origins, University of Southern Denmark,Campusvej 55, 5230 Odense M, Denmark}
\affiliation{INFN - Sezione di Bologna, via Irnerio 46, 40126 Bologna, Italy}

\begin{abstract}
\vspace{3mm}

Scalar field theories with $\mathbb{Z}_{2}$-symmetry are the traditional playground of critical phenomena. In this work these models are studied using functional renormalization group (FRG) equations at order $\partial^2$ of the derivative expansion and, differently from previous approaches, the spike plot technique is employed to find the relative scaling solutions in two and three dimensions.
The anomalous dimension of the first few universality classes in $d=2$ is given and the phase structure predicted by conformal field theory is recovered (without the imposition of conformal invariance), while in $d=3$ a refined view of the standard Wilson-Fisher fixed point is found. 
Our study enlightens the strength of shooting techniques in studying FRG equations, suggesting them as candidates to investigate strongly non-perturbative theories even in more complex cases.

\end{abstract}

\pacs{05.10.Cc,05.30.Rt}
\maketitle

\renewcommand{\thefootnote}{\arabic{footnote}}
\setcounter{footnote}{0}

\section{Introduction}

Since the discovery of the phenomenon of {\it universality} and its explanation in terms of the renormalization group (RG) \cite{Wilson75}, one of the main goals of statistical and quantum field theory has been 
the classification of all universality classes, i.e. the determination of fixed points of the RG flow. 
Even if the phase diagram of  scalar field theories has been subject to intense investigations over the decades, we still lack a complete picture of theory space even in the simplest cases, as that of single component scalar theories in two or three dimensions.
This is not a surprise since the problem is inherently non-perturbative, but what is also missing is an easy way to render and visualize the complex {\it landscape} of theory space, since this is generally an infinite functional space.

In recent years the functional renormalization group (FRG)  \cite{Wetterich93, Morris94}  has shown its versatility and strength as a non-perturbative RG technique in many applications \cite{Berges2002,Schwenk2012,Kopietz2010}. 
In this approach the traditional RG is extended to work in the functional space of the effective average action (EAA), i.e. the scale dependent generator of the one particle irreducible vertexes of the theory, allowing to pursue new kind of non-perturbative expansions and to overcome the traditional limitations of diagrammatic techniques.

In this paper we use the flow equations for the EAA  at order $O(\partial^2)$ of the derivative expansion to investigate scaling solutions (i.e. functional RG fixed points) of single component scalar theories with $\mathbb{Z}_2$-symmetry in two and three dimensions.
Our approach generalizes the spike plot technique already employed to solve the local potential approximation (LPA) in \cite{Morris1996,codello12, codello13, codello14} and refines the $O(\partial^2)$ study \cite{morris95} where a power like regulator and numerical relaxation methods were employed. For related studies, but from a Polchinski equation perspective, see \cite{pol} and reference therein; for a proper time RG study see \cite{alfio}.

\section{Derivative expansion\label{secDE}}

The derivative expansion is an approximation scheme where the EAA is expanded
in powers of the field spatial derivatives \cite{Wetterich93, Morris94, Canet02, LitimZappala}.
This scheme is usually employed for matter field theories on flat space,
where it becomes a series expansion in powers of the momentum.
The derivative expansion has been very successful in drawing phase diagrams and computing accurate universal quantities, especially critical exponents.

If one considers a one component scalar field, in $d$-dimensional flat space, with
a $\mathbb{Z}_{2}$-symmetry, the derivative expansion for the EAA
to order $O(\partial^2)$ reads
\begin{align}
\Gamma_{k}[\phi] = \int d^{d}x\left\{V_{k}(\phi)+\frac{1}{2}Z_{k}(\phi)(\partial\phi)^{2}\right\} +O(\partial^{4})\,.\label{DE}
\end{align}
The effective potential $V_{k}(\phi)$ and the wave-function renormalization
function $Z_{k}(\phi)$ are arbitrary functions of the field $\phi$. 
The derivative expansion has been carried to higher order in $d=3$ \cite{Canet02}, where a beta function study was performed.
The flow equations for these functions are derived from the exact RG equation satisfied by the EAA \cite{Wetterich93, Morris94},
\begin{equation}
\partial_{t}\Gamma_{k}[\phi]= \frac{1}{2} \textrm{Tr} \left(\Gamma^{(2)}_k[\phi]+R_{k} \right)^{-1} \partial_{t}R_{k} \,.
\label{WEQ}
\end{equation}
For translational invariant systems the trace in latter equation represents a momentum integral and the cutoff function $R_k$ is a momentum dependent mass term introduced into the effective action in order to freeze the low energy excitations responsible for the appearance of infrared (IR) divergences. 
The effective action explicitly depends on a scale $k$ and the renormalization "time" is defined as
$t=\log(k/k_{0})$, where $k_{0}$ is an arbitrary reference scale.

After inserting the ansatz \eqref{DE} into the flow equation \eqref{WEQ} and performing the appropriate projection one can derive the {\it beta functionals} $\beta_V$ and $\beta_Z$ for the running effective potential and field dependent wave function renormalization function
\begin{equation}
\partial_t V_k = \beta_V^d (V''_k,Z_k)\qquad \qquad \partial_t Z_k = \beta_Z^d(V''_k,V'''_k,Z_k,Z_k',Z_k'')\,.
\label{BF}
\end{equation}
The explicit form of the beta functionals can be obtained in arbitrary dimension and for arbitrary cutoff functions.
The beta functional for the effective potential follows directly by evaluating \eqref{WEQ} at a constant field configuration and reads
\begin{equation}
\beta_V^d=\frac{1}{(4\pi)^{d\over 2}} \frac{1}{2} Q_{\frac{d}{2}}[G_k\partial_{t}R_{k}]\label{V_flow}\,.
\end{equation}
In \eqref{V_flow} the regularized propagator at the constant
field configuration $\phi$ is defined as
\begin{equation}
G_{k}(x,\phi)=\frac{1}{Z_{k}(\phi)\, x+V_{k}''(\phi)+ R_k(x)}\,,\label{EAA_23.1}
\end{equation}
while the $Q$-functionals are defined as
\begin{equation}
Q_n[f] \equiv \frac{1}{\Gamma(n)} \int^\infty_0 {\rm d} x \, x^{n-1}f(x) \label{Q}\,.
\end{equation}
Deriving the flow equation for the wave-function
renormalization function is more involved. Taking the second functional derivative of \eqref{WEQ} with respect to the fields
it is possible to write down the flow equation for the two-point
function of the EAA in momentum space. Introducing on the rhs of this equation the vertices of
the EAA \eqref{DE} and extracting the
$O(p^{2})$ terms one obtains, after some algebra, the following beta functional
for the wave-function renormalization function
\begin{eqnarray}
\label{Z_flow}
\beta^d_Z&=&\frac{(V_k''')^{2}}{(4\pi)^{d\over 2}}\left\{Q_{\frac{d}{2}}[G_k^{2}G_k'\partial_{t}R_{k}]
+Q_{\frac{d}{2}+1}[G_k^{2}G_k''\partial_{t}R_{k}]\right\}\nonumber\\
&+&\frac{(Z_k')^{2}}{(4\pi)^{d\over 2}}\left\{\frac{2d+1}{2}Q_{\frac{d}{2}+1}[G_k^{3}\partial_{t}R_{k}]
\right. \nonumber\\ &&\qquad\quad\left.
+\frac{(d+2)(d+4)}{4} \left(Q_{\frac{d}{2}+2}[G_k^{2}G_k'\partial_{t}R_{k}]+Q_{\frac{d}{2}+3}[G_k^{2}G_k''\partial_{t}R_{k}]\right)\right\}\nonumber\\
&+&\frac{V_k''' Z_k' }{(4\pi)^{d\over 2}}\left\{2\,Q_{\frac{d}{2}}[G_k^{3}\partial_{t}R_{k}]
+(d+2)\left( Q_{\frac{d}{2}+1}[G_k^{2}G_k'\partial_{t}R_{k}]+Q_{\frac{d}{2}+2}[G_k^{2}G_k''\partial_{t}R_{k}]\right)\right\}\nonumber\\
&+&\frac{Z_k''}{(4\pi)^{d\over 2}} \left\{-\frac{1}{2} Q_{\frac{d}{2}}[G_k^{2}\partial_{t}R_{k}]\right\}\,.
\end{eqnarray}
Equations \eqref{V_flow} and \eqref{Z_flow} represent the flow equations
for $V_{k}(\phi)$ and $Z_{k}(\phi)$ for general
cutoff function at $O(\partial^2)$ of the derivative expansion.
Once an appropriate cutoff shape has been chosen, the integrals
in \eqref{V_flow} and \eqref{Z_flow} can be performed. In
this way one obtains a system of partial differential equations for
$V_{k}(\phi)$ and $Z_{k}(\phi)$ in the variables
$k$ and $\phi$.
Finally the flow equations are obtained introducing the dimensionless variables.
The dimensionless quantities are defined as 
\begin{align}
\phi =  Z_k^{-\frac{1}{2}}\, k^{\left(\frac{d}{2}-1\right)}\varphi \qquad\qquad  V_k(\phi) = k^d\, v(\varphi)\qquad\qquad Z_k(\phi) = Z_k\, \zeta (\varphi) \,,
\end{align}
from which we drive the following relations by $t$-differentiation
\begin{equation*}
\beta_v  = - d \, v + \frac{d-2+\eta}{2} \,\varphi \,v'  + k^{-d} \beta_V
\end{equation*}
\begin{equation}
\beta_\zeta = \eta\, \zeta + \frac{d-2+\eta}{2} \,\varphi\, \zeta' + Z_k^{-1} \beta_Z \,,
\label{dimlessDE}
\end{equation}
where $\beta_v \equiv \partial_t v$ and $\beta_\zeta \equiv \partial_t \zeta$.
The anomalous dimension of the scalar field in \eqref{dimlessDE}
is defined by $\eta=-\partial_{t}\log Z_{k}(0)$  \cite{Berges2002}.

To obtain explicit expressions for the beta functional we need to specify the cutoff function $R_k$.
In terms of the linear cutoff\footnote{The linear cutoff is believed to be the optimal choice at low order in derivative expansion \cite{Litim2000}. Indeed the results for the critical exponents found in this case are more accurate than the ones obtained with power law cutoff \cite{Morris94,morris95}, as discussed in the following.} introduced in \cite{Litim2000} 
\begin{equation}
R_k(x) = Z_k (k^2-x)\theta (k^2-x) \,,
\label{litim}
\end{equation}
the $Q$-functionals in \eqref{V_flow} and \eqref{Z_flow} can all be computed analytically and they can be reduced to a single threshold integral, a hypergeometric function (see the Appendix), which in integer dimension simplifies even further to elementary expressions. Explicit expressions for $\beta_v$ and $\beta_\zeta$ can then be written in two and three dimensions, as shown in the Appendix.

\section{Scaling solutions}
In this formalism the scaling solutions for the effective actions appear as FRG fixed points and they are determined by solving the coupled system of ordinary differential equations
\begin{equation}
\beta_v =0\qquad\qquad\qquad \beta_\zeta=0\,.\label{EAA_25}
\end{equation}
For parity reasons we expect the following boundary conditions for the effective potential and the wave-function renormalization functional evaluated at the origin,
\begin{align}
v'(0)&=0\label{cond_1}\\
v''(0)&=\sigma\label{cond_2}\\
\zeta(0)&=1\label{cond_3}\\
\zeta'(0)&=0\,.\label{cond_4}
\end{align}
Conditions \eqref{cond_1} and \eqref{cond_4} are a direct consequence of the $\mathbb{Z}_{2}$-symmetry. The condition \eqref{cond_4} is obtained absorbing a factor $Z_k(0)$ into the field redefinition.
Thus the only unspecified condition remaining is  the $v''(0)$ value. In equation \eqref{cond_2} $\sigma$ and $\eta$ are real values to be determined by requiring the global existence of the scaling solutions, condition that, as we will see, reduces to a finite set the allowed functional fixed points. 

The main goal of our paper is to show that it is possible to find solutions of the system \eqref{EAA_25} extending the simple spike plot technique used in LPA analysis \cite{Morris1996,codello12}. This method has diverse advantages: first of all it does not require the solution of the flow equations in function of the renormalization time $t$; second it solves the fixed point equations in their full functional form, without relying on truncations \cite{Canet02} (this property is necessary to be consistent with the Mermin-Wagner theorem \cite{Defenu15}); third it does not necessitate any external input as in \cite{Morris94, morris95}, where the relaxation method was used starting from the exact spherical model solution.
For these reasons the present method is the most suited to uncover actual non-perturbative universality classes, which can not be investigated by means of standard approaches \cite{shunsuke17}.

Let us describe the procedure in more details, we solve equations \eqref{EAA_25} for different values of $\sigma$ and $\eta$. For any arbitrary point in the $(\sigma,\eta)$ plane the solution will become singular at a finite value of the field \cite{zanussowipf}, we call this finite value $\varphi_{\rm{sing}}$, and global scaling solutions correspond to those points in which $\varphi_{\rm{sing}}$ shows a "spike" behaviour. Thus physical fixed points in the $(\sigma,\eta)$ plane can be determined by a numerical analysis of the function $\varphi_{\rm{sing}}$.

For the purpose of extending the resulting scaling solutions $v$ and $\zeta$ beyond $\varphi_{\rm{sing}}$
the analysis is complemented by the asymptotic behaviours computed using the large field solutions of equations \eqref{EAA_25}
\begin{align}
\label{Eq4}
\begin{cases}
  v(\varphi)&\sim\, v_{0}\,\varphi^{\frac{2\,d}{d-2+\eta}}\\
  \zeta(\varphi)&\sim\, \zeta_{0}\,\varphi^{-\frac{2\eta}{d-2+\eta}}
  \end{cases}\,\,\,\, {\rm for} \,\,\,\, \varphi  \gg 1\,.
\end{align}
Once the location of a fixed point in the $(\sigma,\eta)$ plane has been determined, the coefficients $(v_{0},\zeta_{0})$ can be evaluated by imposing continuity of the numerical solutions for $v$ and $\zeta$ with the large field expansion in Eq, \eqref{Eq4} at some value $\varphi_{\rm{max}}<\varphi_{\rm{sing}}$. 

We will show in subsection \ref{3d} that
in $d=3$ there is only one scaling solution to the system \eqref{EAA_25}, which
corresponds to the so-called Wilson-Fisher
fixed point and describes the {\tt Ising} $d=3$
universality class.
It is in $d=2$, where every perturbative approach fails
to describe correctly the various universality classes, first constructed
exactly using conformal field theory (CFT) methods \cite{Mussardo10},
that the system \eqref{EAA_25}) reveals its non-perturbative potentialities.
It was shown in \cite{Morris94} that the power-law cutoff version of the system \eqref{EAA_25} 
has scaling solutions in one-to-one correspondence
with the minimal models known from CFT.
In subsection \ref{2d} we will see that the same picture emerges also when the linear cutoff and the spike plot method are employed.

\subsection{The three dimensional case\label{3d}}

\begin{figure}
\centering
 \subfigure[\;Contour plot of $\varphi_{\rm{sing}}(\sigma,\eta)$ in $d=3$]
   {
   \includegraphics[scale=.25]{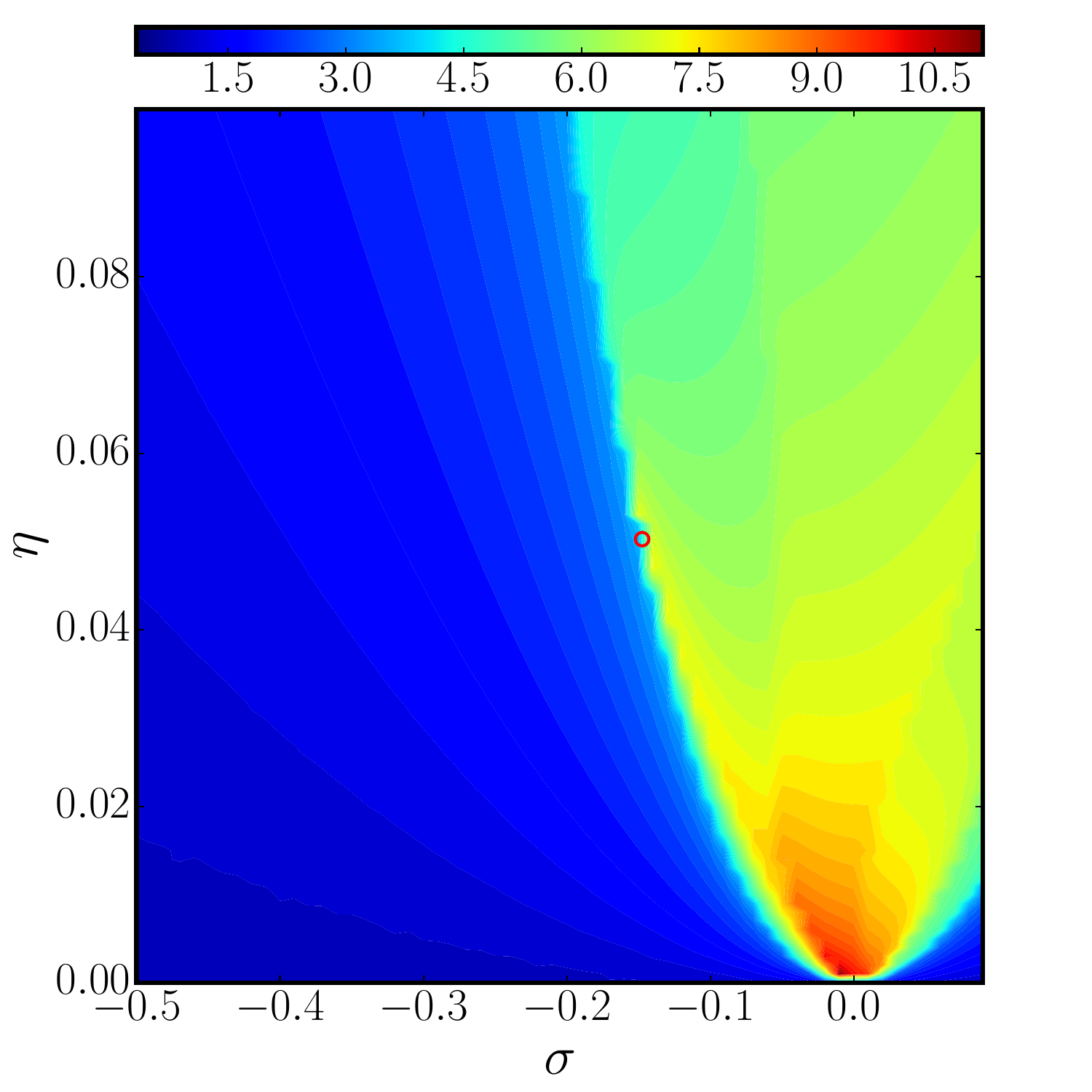}
   }
\hspace{5mm}
\subfigure[\;Spike-plot along the chain of maxima in the contour plot of panel (a).]
   {
\includegraphics[scale=0.25]{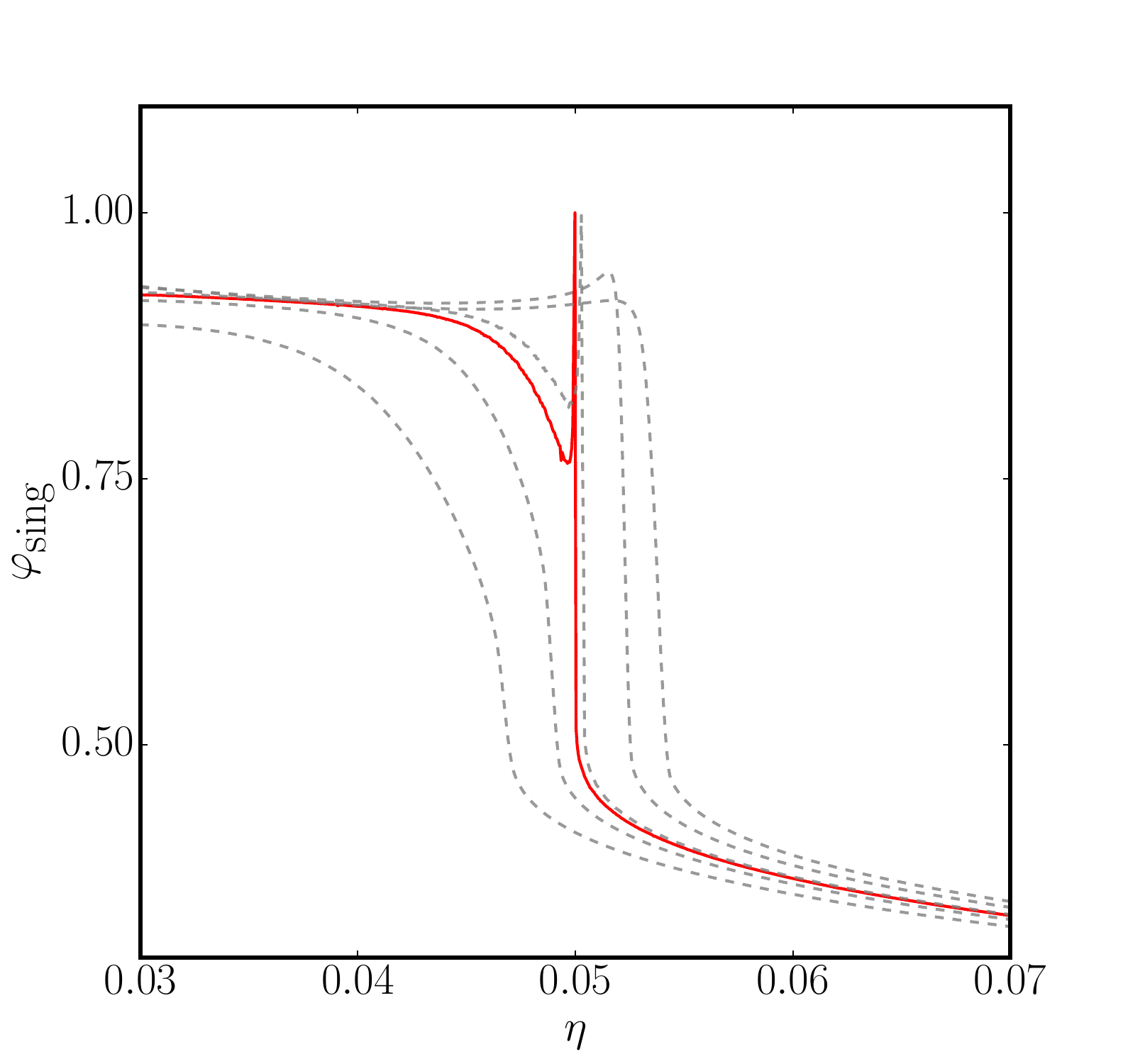}
}
 \caption{The contour plot of $\varphi_{\rm{sing}}$ in the $(\sigma,\eta)$ planes has two spikes: the one located at $(0,0)$ represents the {\tt Gaussian} theory. The other one indicates the {\tt Ising} universality class, which occurs at a finite positive value of the anomalous dimension and for a negative dimensionless mass. In panel (b) the spike-plot is performed along the maxima chain of the contour plot of panel (a) with different choices for the fit parameters (gray dashed lines). The optimal choice (red solid line) is the one which maximizes the spike height, see the main text. \label{Fig1}}
 \end{figure}

First of all we consider the $d=3$ case, which is the less involved since only the standard Wilson-Fisher fixed point exists.
The equations for the $\beta$-functions in the linear regulator case, \eqref{flow_V_d=3_litim} and \eqref{flow_Z_d=3_litim} from the Appendix, are set to zero and solved numerically, with a fourth order Runge-Kutta approach, for several points in the $(\sigma,\eta)$ plane. 
The value at which the numerical solver interrupts integration $\varphi_{\rm{sing}}$ is represented as a contourplot in the plane of the initial conditions $(\sigma,\eta)$ in
Fig.~\ref{Fig1} panel (a). 

Apart for the maximum in the origin, which represents the free {\tt Gaussian} universality class, we see a family of maxima arising along a curve which separates the dark blue region of the severely ill defined solutions from the light green region of intermediate solutions.
Only a small  portion of the plane has a non-trivial support for the $\varphi_{\rm{sing}}$ function, which is identically zero for all $\sigma<-0.2$.
Moreover the equations in the linear cutoff case develop a divergence for $\sigma=-1$ often called spinodal instability. It is not clear wether this divergence, which is IR attractive in all directions, is the physical IR fixed point or rather an artifact of the linear regulator scheme. However it has been demonstrated that it is possible to recover information over the universal quantities studying the scaling of the solutions close to this instability \cite{nagy12}. 

The landscape in Fig.~\ref{Fig1} can seem surprising due to the presence of several maxima rather than an isolated one, indicating the \texttt{Ising} universality class. However a closer inspection reveals that the high of the peaks is non monotonous, leading to the appearance of a prominent maximum at finite $\eta$ and $\sigma$ values. This prominent maximum represents the signature of the Wilson-Fisher fixed point. The connection of the \texttt{Ising} and \texttt{Gaussian} universalities through an extended chain is easily justified if one considers that the spike-plot technique furnishes a solution even in the lower truncation scheme, where only the potential equation is considered. Thus it appears that for each value of $\eta$ it exists a maximum as a function of $\sigma$ which represents the best approximation for the $v(\varphi)$ and $\zeta(\varphi)$ functions at that particular value of the anomalous dimension.

On the other hand due to the finite numerical grid in the $(\sigma,\eta)$ plane, to the complex three dimensional shape of the chain as well as to the finite numerical accuracy in the solution of the flow equations. The peaks chain appears as a sequence of spikes rather than a continuous ridge. This picture is confirmed in the following where we pursue the analysis along several trajectory in the $(\sigma,\eta)$, retrieving the expected scenario.
\begin{figure}
\centering
\subfigure[\;Fixed point potential $v(\varphi)$ in $d=3$]
{
\includegraphics[scale=0.25]{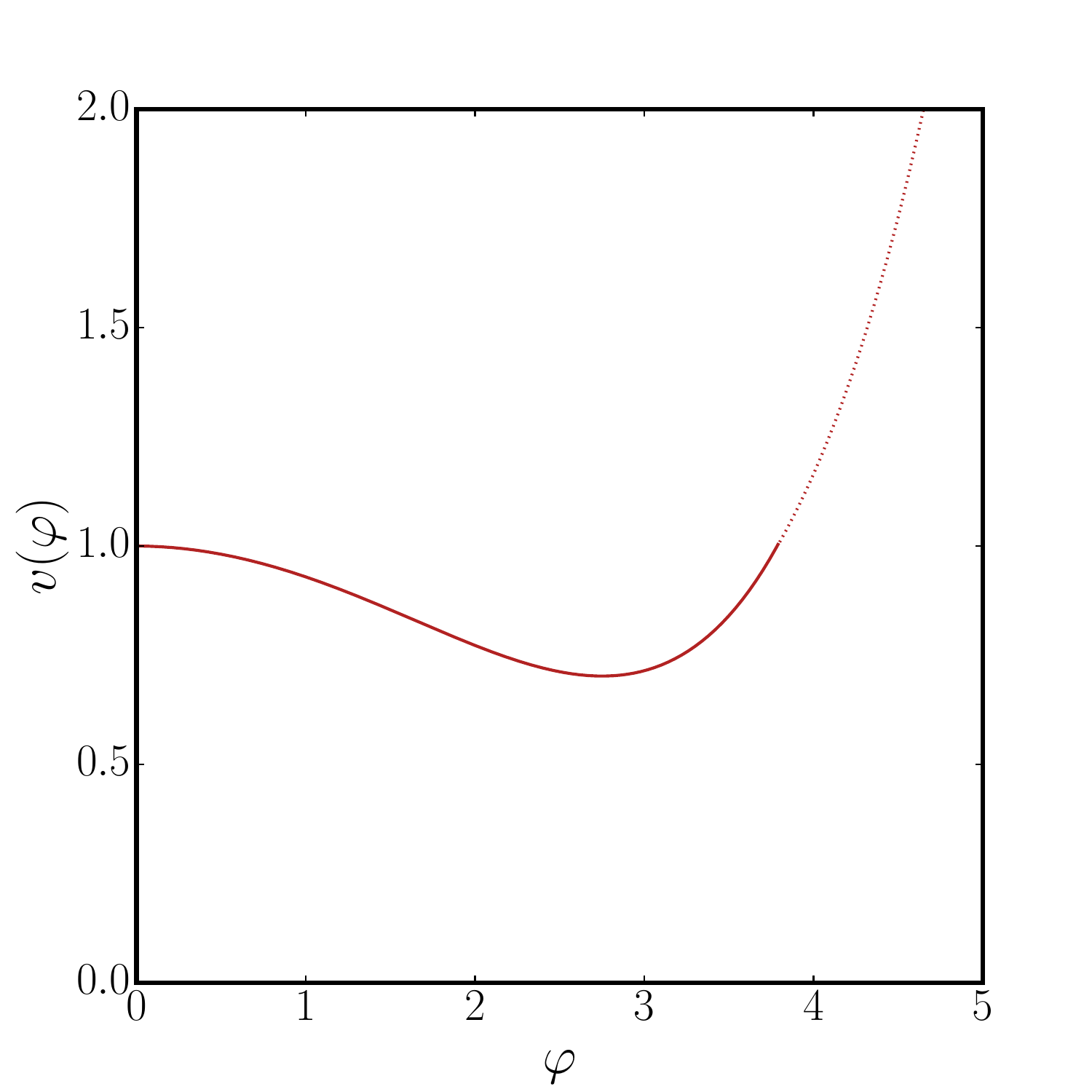}
}
\hspace{5mm}
\subfigure[\;Fixed point wave-function $\zeta(\varphi)$ in $d=3$]
{
\includegraphics[scale=0.25]{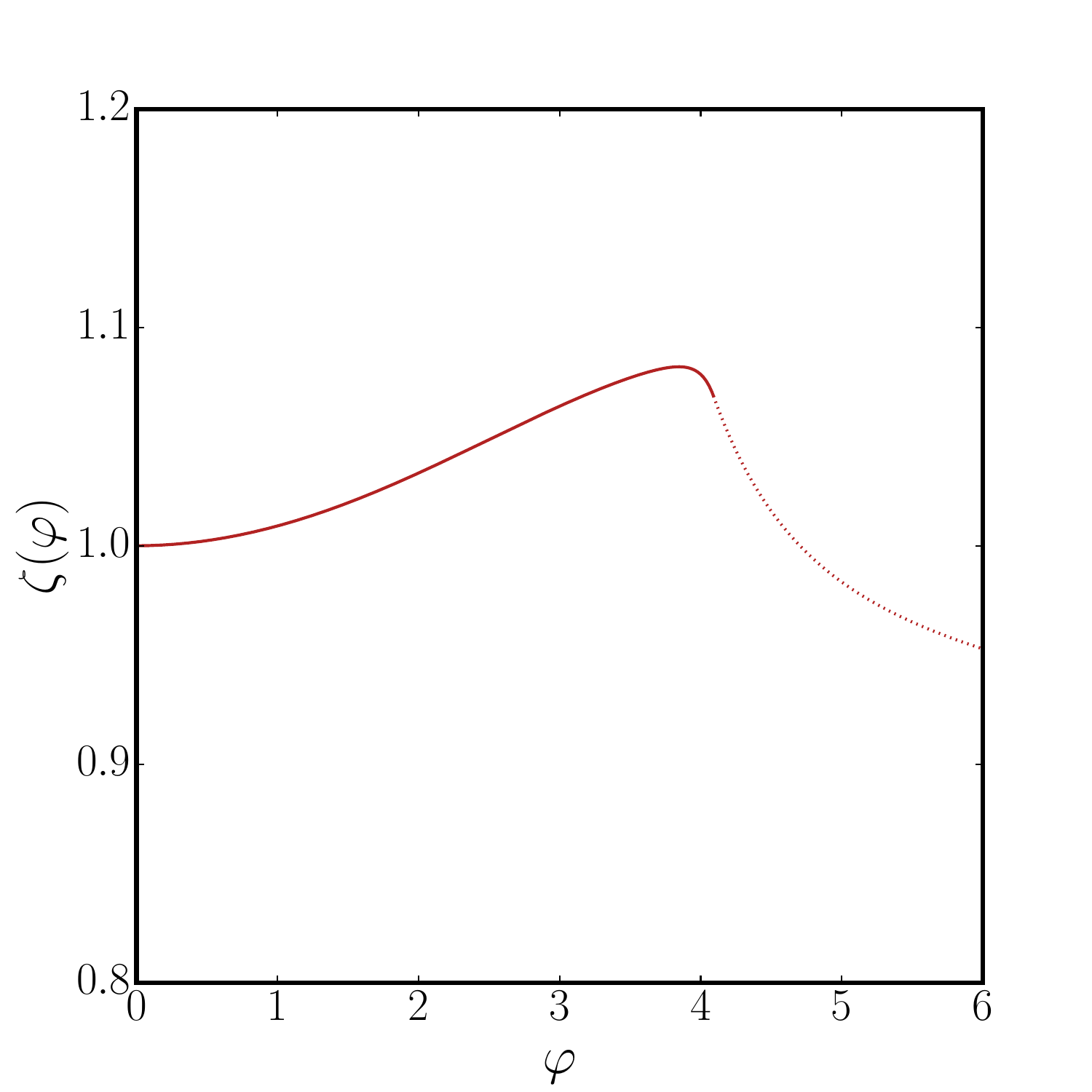}
}
 \caption{In panel (a) and (b), respectively, the solutions for the functions $\zeta(\varphi)$ and $v(\varphi)$ are shown for the Wilson-Fisher fixed point in three dimension. The lines have been obtained imposing continuity between the numerical solution (solid line) and analytic large field behaviour of the scaling solutions \eqref{Eq4} (dotted lines)}. \label{Fig3}
 \end{figure}
As anticipated above a convenient approach to overcome these difficulties and to extract the anomalous dimension of the interacting universality classes is to reduce the problem of finding the maxima of the two dimensional surface $\varphi_{\rm{sing}}(\sigma,\eta)$ to the simpler one of locating the maximum in a single one dimensional chain. This is possible using the results in panels (a) of Figs. \ref{Fig1} and \ref{Fig2} as a guide.

First of all we fit the locations of the maxima of a single chain in the ($\sigma$,$\eta$) plane with a simple function, depending on the cases  linear or parabolic fit functions are employed. This operation produces an explicit expression for the maxima chain as a function $\sigma(\eta)$ along which we can pursue the standard one dimensional spike plot technique.
Obviously the coefficients of the fit will contain errors due to the finiteness of the grid in the landscapes of panels (a) in figures \ref{Fig1} and \ref{Fig2}. However the best value for the fit is obtained optimizing the coefficient in order to maximize the height of the spike; as it is shown in panel (b) of Fig.~\ref{Fig1} for the three dimensional {\tt Ising} universality. 
The optimization procedure is straightforward since one should allow only for very small variation of the fit coefficients and the one dimensional spike plot computation is extremely fast.

In the three dimensional case the optimization procedure is shown in Fig.~\ref{Fig1} panel (b). The grey dashed curves are the spike plots along different fit representations of the peak chain in panel (a), Fig.~\ref{Fig1}. The equation for the peak chain has been obtained by linear fit around the position of the non-trivial maximum. The various curves (gray dashed lines) show the result for different values of the fit coefficients, the optimized curve, red solid line, gives $\eta=0.0496$ with $\sigma=-0.1468$.
The fixed point potential and wave-function renormalization are shown in Fig.~\ref{Fig3} panel (a) and (b) respectively. In correspondence of the minimum of the fixed point potential a maximum of the wave-function is found. The large field behavior
can not be obtained by numerical integration, since a very precise identification of the correct fixed point values $(\sigma,\eta)$ is necessary to push the integration forward. In Fig.~\ref{Fig3} the large field branch of the two functions has been obtained using large field expansions reported in equation \eqref{Eq4} and imposing continuity with the numerical solutions.

\subsection{The two dimensional case\label{2d}}

In the $d=2$ case scalar $\mathbb{Z}_{2}$-theories have an infinite number of universality classes, leading to a far more complicated landscape in the $(\sigma,\eta)$ plane. These universality classes correspond with the minimal models of CFT and the critical exponents are known exactly.
The importance of reproducing such results in an approximated scheme is not only the methodological one, deriving from the necessity to test our technique in a more complex scenario, but it also lies in the different assumptions necessary to compute such quantities.  Indeed flow equations \eqref{V_flow} and \eqref{Z_flow} have been derived without any additional condition rather than $\mathbb{Z}_{2}$-symmetry. Even in this oversimplified computation scheme and without any additional imposition on the symmetry of the model, such as conformal symmetry, the FRG technique is able to recover all the information on the location and the shape of the solutions, with a good numerical accuracy on the universal quantities.
\begin{figure}[t]
\centering
\subfigure[\;Contour plot of $\varphi_{\rm{sing}}(\sigma,\eta)$ in $d=2$]
{
\includegraphics[scale=0.25]{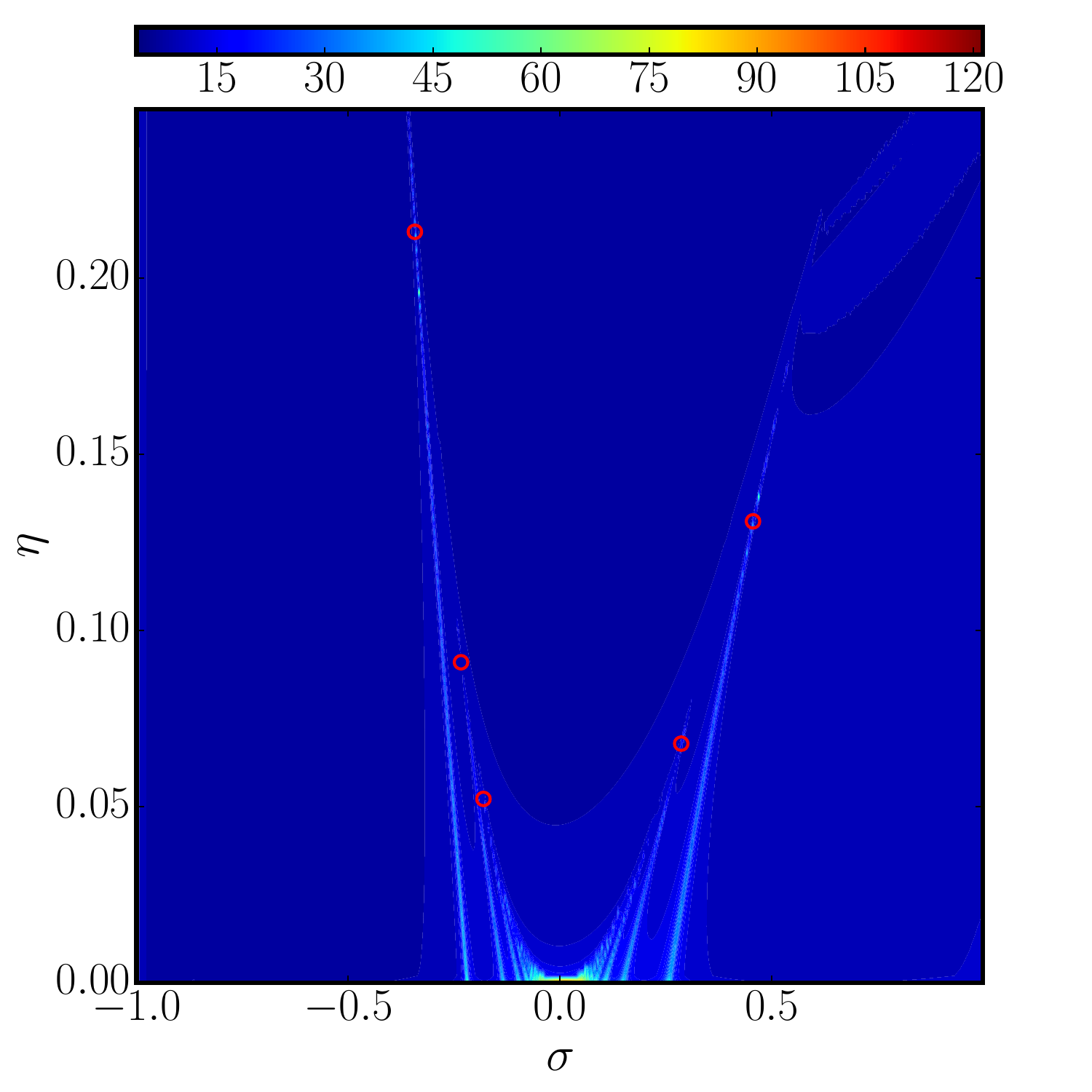}
}
 \hspace{5mm}
\subfigure[\;Spike plots along the chains of panel (a)]
{
\includegraphics[scale=0.3]{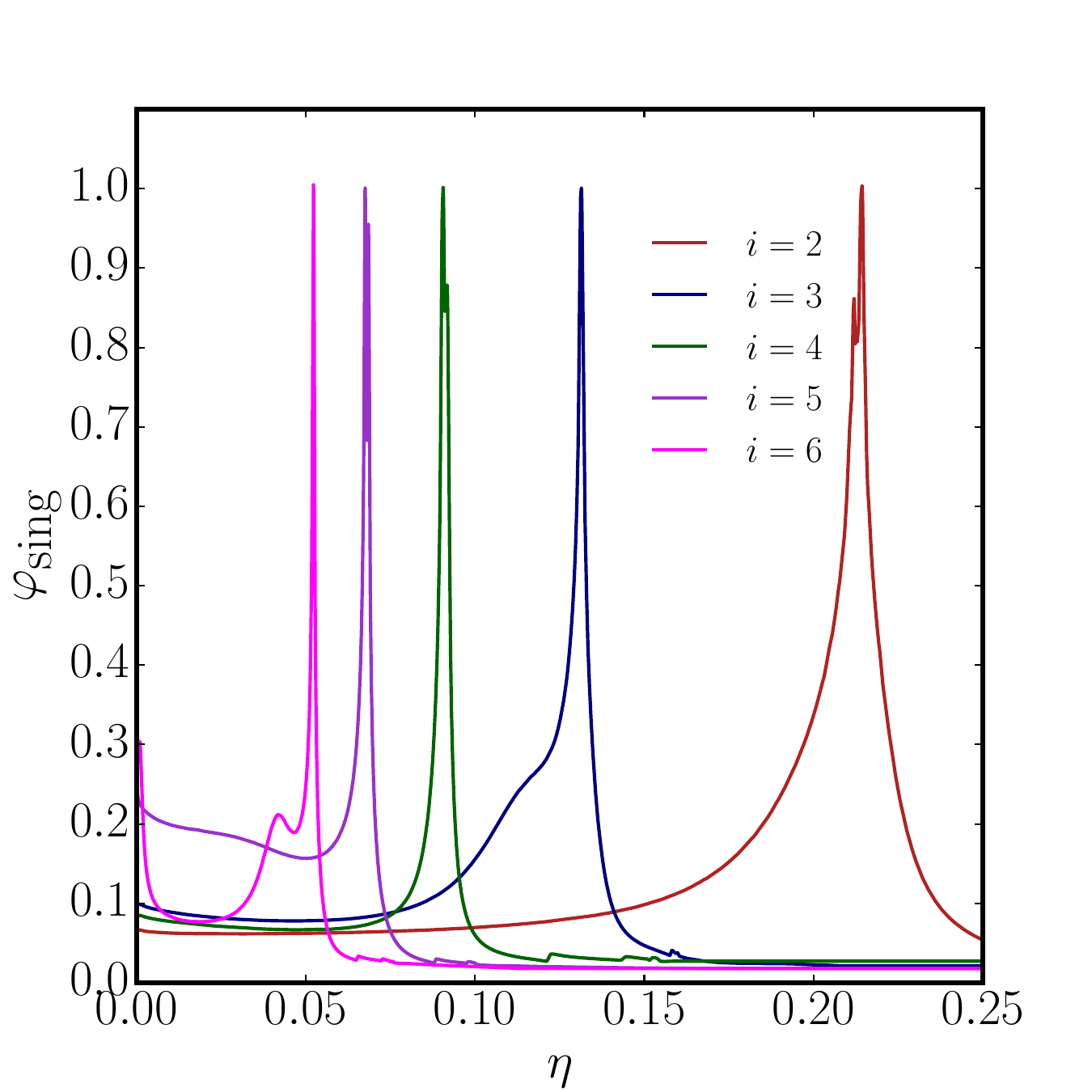}
}
 \caption{In panel (a) spike plot in two dimensions for the linear regulator case. The landscape of scalar quantum field theories at this approximation level is quite complicated. The maxima of $\varphi_{\rm{sing}}$ form mountain chains located on some special curves in the $(\sigma,\eta)$ plane.\label{Fig2}}
 \end{figure}

The result for the landscape of interacting fixed points in two dimensional scalar field theories is reported in Fig.~\ref{Fig2}, in the linear cutoff scheme. Identifying correctly the exact value of the anomalous dimension for any universality class requires some care. Indeed, similarly to the $d=3$ case in Fig.~\ref{Fig1}, the maxima of $\varphi_{\rm{sing}}$ are disposed on special lines of the $(\sigma,\eta)$ plane, forming peak chains.
In panel (a) of Fig.~\ref{Fig3} the {\tt Gaussian} universality class appears as a infinitely tall spike located at $\sigma=\eta=0$. The two peaks chains at the extrema of the origin are the longest in the $\eta$ direction and they represent the Wilson-Fisher and tricritical universalities which have the largest anomalous dimension values. Both these peak chains, going from $\eta=0$ to $\eta\simeq1$, have a maxima at a finite value of the anomalous dimension which roughly corresponds to the expected one.
\begin{figure}
\centering
\subfigure[\;Anomalous dimensions of the first six {\tt Multi-critical} universality classes in $d=2$]
{
\includegraphics[scale=0.3]{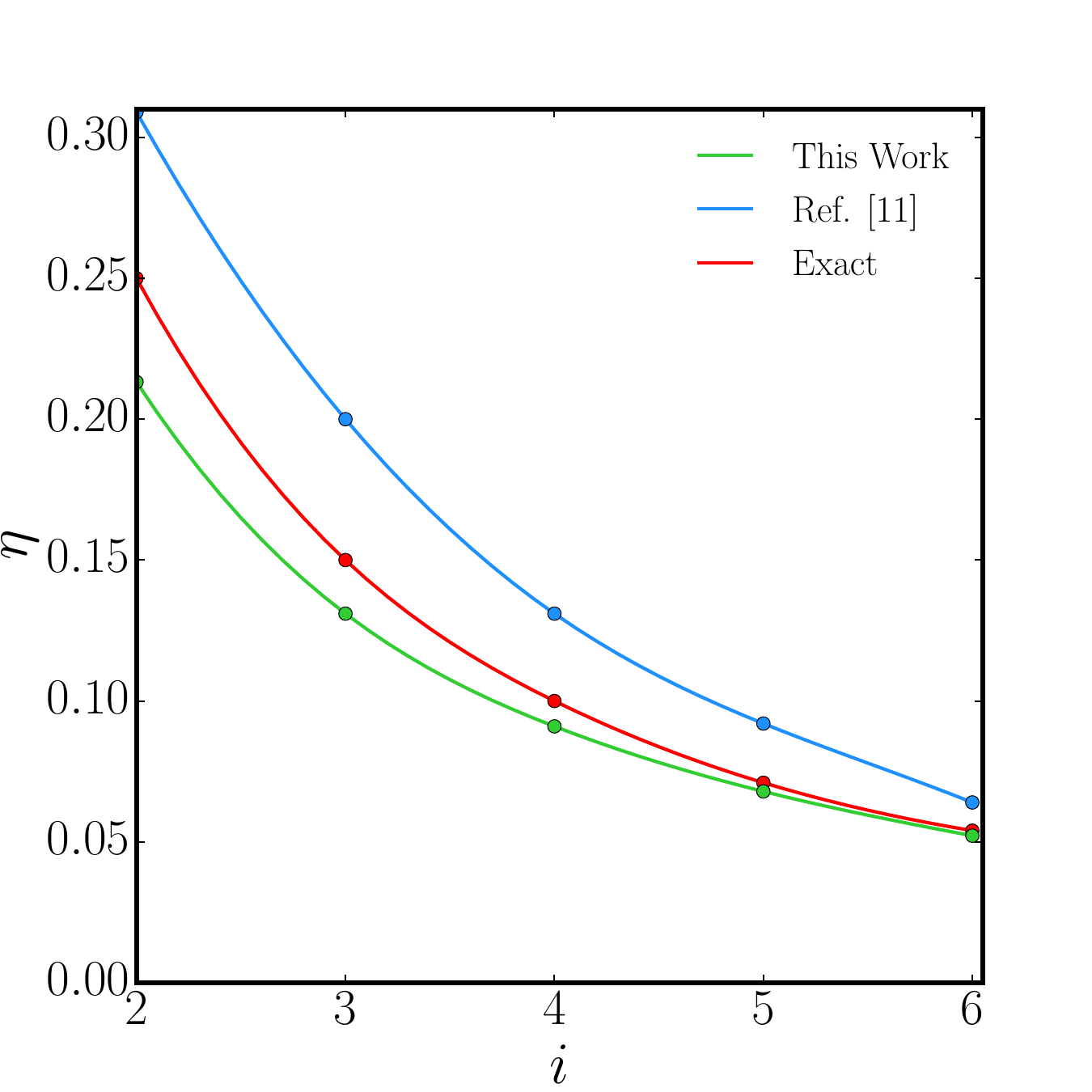}
}
\hspace{5mm}
\subfigure[\;Fixed point solutions for the functions $v(\varphi)$ and $\zeta(\varphi)$ in $d=2$]
{
\includegraphics[scale=0.27]{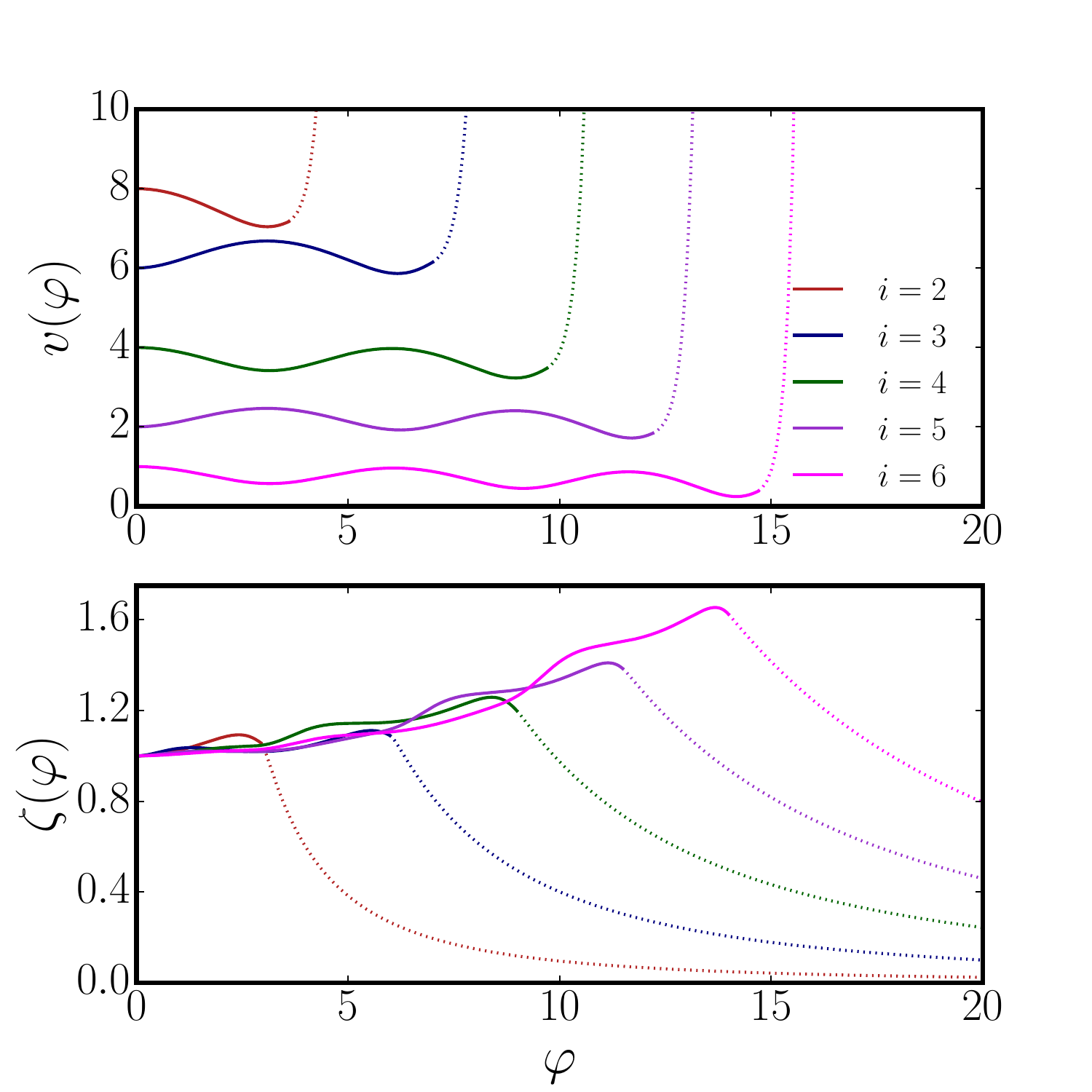}
}
\caption{In panel (a) we show the anomalous dimensions as a function of the critical index $i$ for the linear (green line) and power law (blue line) cutoff case, compared to the exact CFT results (red line). In panel (b) the solutions for the functions $\zeta(\varphi)$ and $v(\varphi)$ are shown for the first six multi-critical universalities. Each line has been obtained imposing continuity between the numerical solution (solid line) and analytic large field behaviour of the scaling solutions \eqref{Eq4} (dotted lines). \label{Fig4}}
\end{figure}
Moreover we have infinitely many other peak chains starting at $\eta=0$. These chains accumulate in the origin, becoming shorter and shorter, as expected since they represent high order universalities.  

As for the three dimensional case, the finite numerical grid in the $(\sigma,\eta)$ plane does not allow for a precise location of the tallest maximum along each chain. Increasing the precision of the grid will need an exponential growing computation time due to the necessity to increase accordingly the precision of the numerical solution of the differential equations. Once again, in order to maintain a low computational cost, we consider a numerical fit for each peak chain along which we purse a one dimensional spike plot procedure analogue to the LPA case.
Optimizing the fit coefficients to maximize the height of the peak in every chain of the contour-plot in $d=2$, Fig.~\ref{Fig2} panel (a), we obtain a one dimensional spike plot for every universality class from the standard Wilson-Fisher case $i=2$ to the esacritical universality $i=6$ as it is shown in Fig.~\ref{Fig2} panel (b). The values, at which the singularities of $\varphi_{\rm{sing}}$ occur, are in agreement with the expected values for the anomalous dimensions.
During the optimization procedure of the fit parameters it occurs that two minima are found in the line, proving the complex three dimensional structure of the peaks in the ($\sigma$,$\eta$) plane. In any case the highest divergence is always obtained for a single peak, confirming that the two peaks structure occurs only when the fit line does not cross the spike at its center. The distance between the two peaks can be easily reduced until a precision $\leq 5\%$ is obtained in the $\eta$ values, see Fig.~\ref{Fig2} panel (b).

In Fig. \ref{Fig4}, panel (a), the anomalous dimension values obtained using the spike plot technique are shown. The results of this work (green line) are compared to the ones found in \cite{morris95} (blue line) and to the exact results of CFT solutions (red line). The curve obtained here is more precise than the results found in the power law cutoff scheme of \cite{morris95}, which confirms the expected better performance of the linear cutoff scheme.
It should be also noted that, as expected, the precision of FRG truncation scheme increases lowering the anomalous dimension values, with the esacritical value for the $\eta$ exponent reproduced up to $99\%$ even in this rather simple approximation.
The fact that the precision of FRG truncation scheme increases lowering the anomalous dimension values suggests that a truncation based on the real fixed point dimension of the operators included in the EAA might be better than the derivative expansion from the quantitative point of view. While in $d=2$ for $\eta=0$ the expansion in terms of the operator dimensions and derivative expansion coincides as $\eta \to 0$ and thus, as $\eta$ grows, they start diverging. In $d=3$ the situation is very different and may explain the less precise numerical success of the derivative expansion in this dimension.

In panel (b) of Fig.~\ref{Fig4} we show the potential $v(\varphi)$ and the field dependent wave function renormalization $\zeta(\varphi)$ for the first six universality classes in $d=2$. The solutions are shown only for positive values of the field $\varphi$, since the other branch can be simply obtained using reflection symmetry. Each potential shows a number of minima $i$, as indicated by its criticality order, and the corresponding wave function renormalization has a relative maximum in correspondence of each minimum position.
The solutions shown in Fig.~\ref{Fig4} panel (b) have been obtained solving equations \eqref{V_flow} and \eqref{Z_flow} with the values of $\eta$ and $\sigma$ found using the spike plot technique described in Fig.~\ref{Fig1} panel (b).

\subsection{Regulator dependence}

As it should be understood from the above investigations, the efficiency of our spike plot technique crucially depends on the structure of the three dimensional surface $\varphi_{\rm{sing}}(\sigma,\eta)$, which has non-trivial shape only on a finite number of quasi-two dimensional manifolds, i. e. the lines in Fig. \ref{Fig2}. It is then necessary to test wether this simplified structure is just an artifact of the linear cutoff function or if it is a general result valid for equations \eqref{V_flow} and \eqref{Z_flow} independently from the particular form assumed by the $Q$-functionals.
\begin{figure}
\centering
\subfigure[\;Contour plot of $\varphi_{\rm{sing}}(\sigma,\eta)$ in $d=2$ for the power law regulator]
{\includegraphics[scale=0.28]{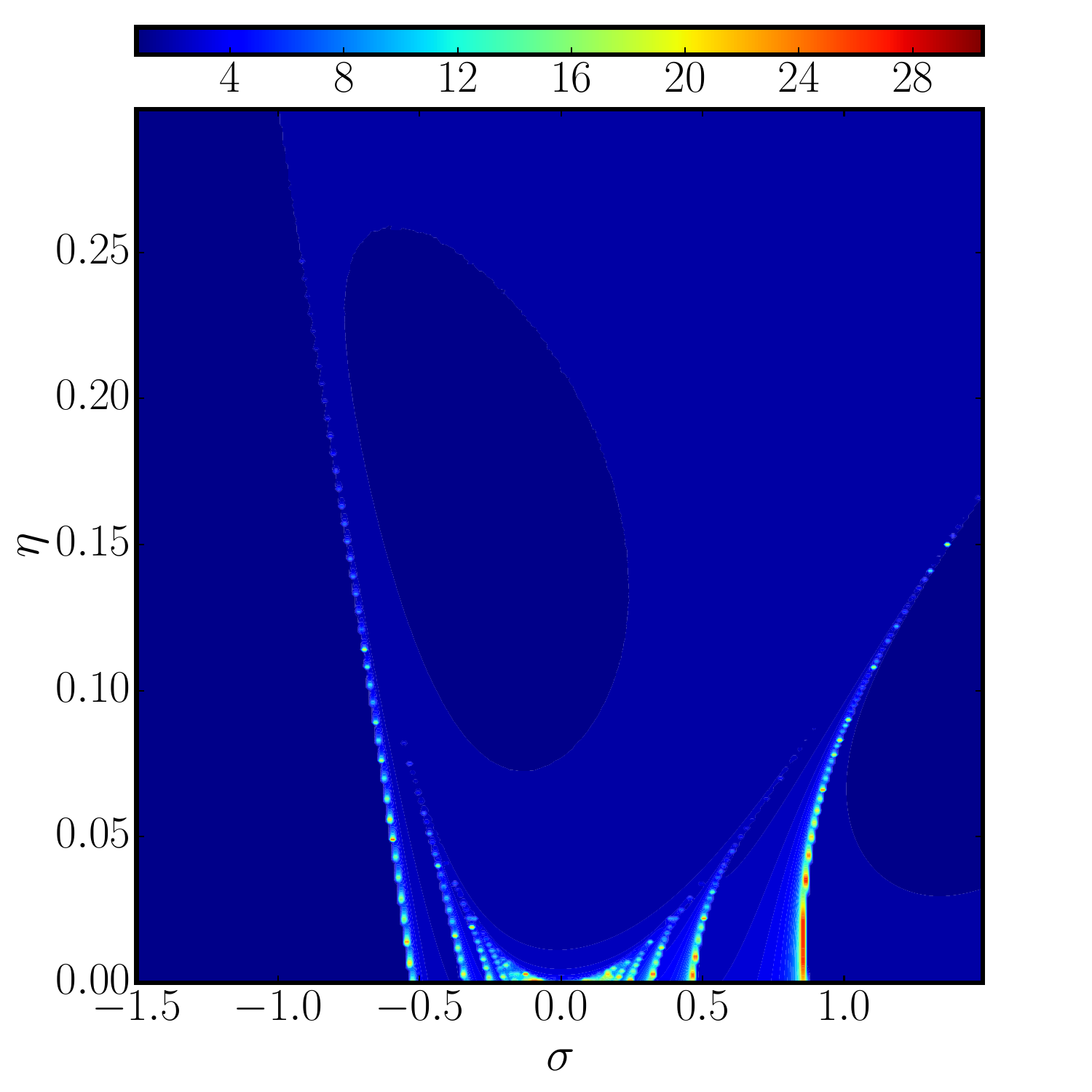}}
\hspace{5mm}
\subfigure[\;Contour plot of $\varphi_{\rm{sing}}(\sigma,\eta)$ in $d=3$ for the power law regulator]
{\includegraphics[scale=0.28]{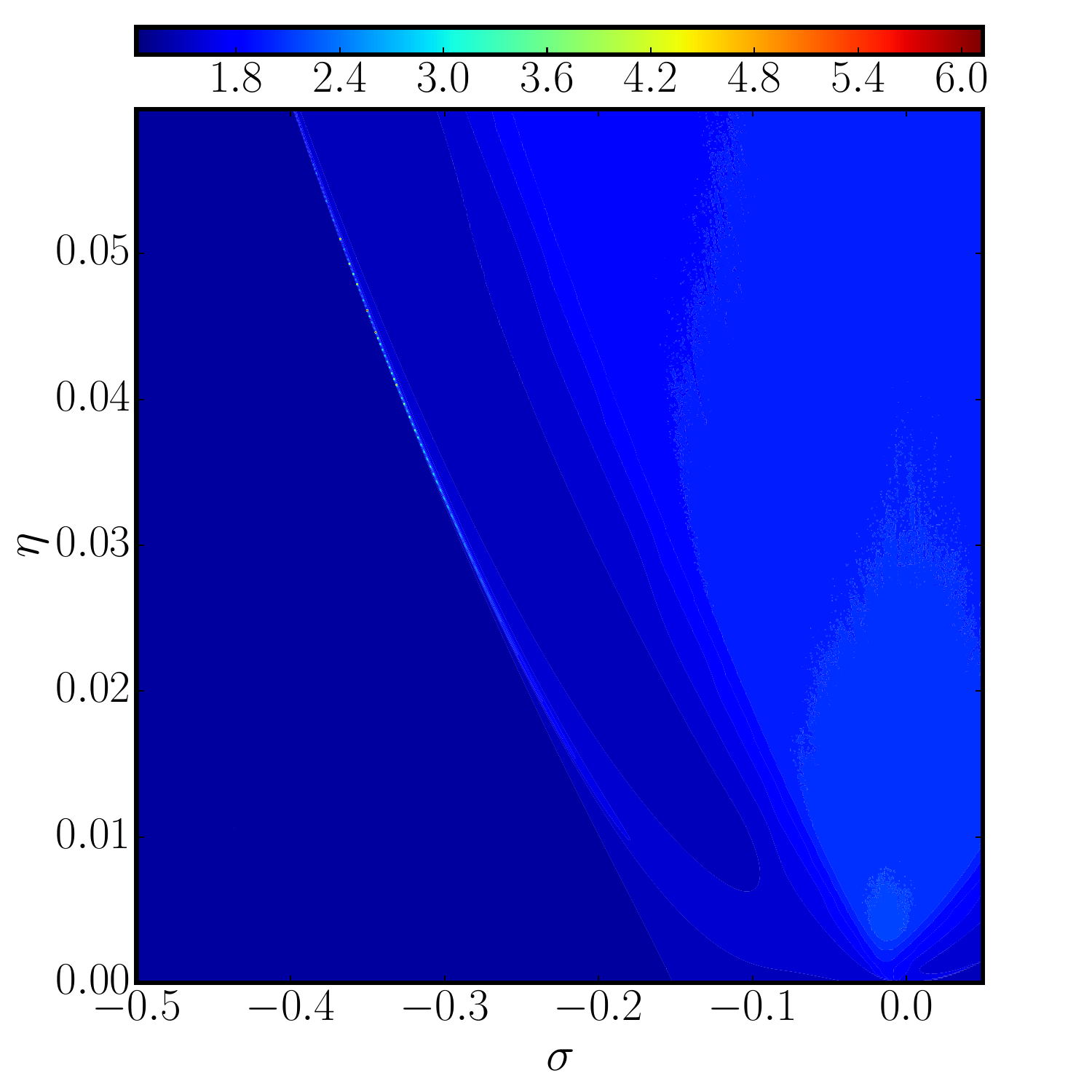}}
\caption{Spike plot in two dimensions for the power law cutoff case. The landscape of scalar quantum field theories at this approximation level is quite complicated. The maxima of $\varphi_{\rm{sing}}$ form mountain chains located on some special curves in the $(\sigma,\eta)$ plane.\label{Fig5}}
\end{figure}

In order to check the stability of our approach we consider another cutoff function to explicitly compute the $Q$-functionals. The most effective choice in this perspective is the power law regulator already adopted in \cite{morris95}
%
$
R_{k}(x)=k^{4} / x\,.
$
%
Indeed such regulator, even if not optimal to compute numerical quantities, produces simple results for the flow equations. Moreover the power law cutoff has a rather peculiar shape, completely different from the linear cutoff employed above. Indeed while the last one is compact with extremely localized derivatives the first one has infinite support and it has finite derivatives at all orders.

We apply the same procedure already considered for the flow equations in the linear regulator case. Both in $d=2$ and $d=3$ we retrieve the expected phase structure, with only one correlated fixed point in the first case and infinitely many solutions in $d=2$, as shown in Fig. \ref{Fig5}.
It is worth noting that in the power law regulator case the height of the peaks in the contour plot is much smaller than in the previous case. However both in 2, Fig. \ref{Fig5} panel (a), and 3, panel (b),
the peak chains are evident, thus demonstrating that the structure of the $\varphi_{\rm{sing}}(\sigma, \eta)$ function, even if influenced by the regulator form, maintains a very small non-trivial support. In both the regulator cases every universality class of the theory appears as a chain of peaks in the $(\sigma, \eta)$ plane.

\section{Conclusion}

After deriving the explicit expressions for the flow equations of the effective potential $V_k$ and the wave-function renormalization function $Z_k$, in two and three dimensions, and with the use of the linear   cutoff, we have shown that it is possible to extend the "spike plot" technique to the study of scaling solution at order $\partial^2$ of the derivative expansion. 
In this approximation theory space is projected to the two dimensional plane parametrized by $\sigma \equiv v''(0)$ and by the anomalous dimension $\eta$.
The spike plot becomes a two dimensional surface that represents the landscape of scalar theories with $\mathbb{Z}_2$-symmetry.
These landscapes are characterized by "peak chains" along which the spikes appear; the shape of these chains can be fitted by a simple linear or quadratic curve $\sigma(\eta)$ and the problem of finding the position of the maxima of these surfaces can be reduced to a one dimensional spike plot similar to those found at the LPA level.

In $d=3$ the landscape is characterized by a single non-trivial spike that represents the {\tt Ising} universality class, while in two dimensions the landscape is much more complex: a whole chain is present for each {\tt Multi-critical} universality class and each CFT minimal model emerges as the most prominent peak of each of these chains. At high multi-criticality the obtained values of the anomalous dimensions are in very good agreement with the exact CFT results.
At the quantitative level, the increasing precision with which the anomalous dimension of the multi-critical fixed points is obtained, for growing multi-criticality index, suggests that an expansion scheme based on the relevancy of the {\it real} operator dimension (operator dimension expansion) of the terms included in the truncation ansatz of the effective average action might be quantitatively precise. This is suggested by the fact that the derivative expansion, in $d=2$ and as $\eta\to 0$, becomes exactly the operator dimension expansion and indeed we obtain better and better values for $\eta$ as the multi-criticality grows.

While the present analysis shows how the spike plot technique can produce a complete and systematic analysis of fixed point solutions of functional flow equations, even beyond the well known LPA case, it would be interesting to disclose the full picture of the critical exponents for scalar field theories, including the correlation length exponent $\nu$, even for fractional dimension beyond two and three where it will be interesting to compare with the recent $\epsilon$-expansion analysis \cite{osborn,bologna,bologna2,bologna3}. However the latter task is hindered by the complexity of the flow equations, even in the case of the linear regulator \eqref{litim}, which leads to substantial numerical difficulties, especially in the study of the stability exponents. 
Then, it would be more convenient to consider different regularization schemes which deliver considerably simpler functional flow equations \cite{Zambelli2015}.  Another interesting perspective is the study of the non-unitary family of fixed points generalizing the LPA analysis \cite{Zambelli2016} or long range interacting field theories, both in the classical \cite{Defenu2015} and quantum case \cite{Defenu2016,Defenu2017} or, finally in presence of global symmetries as $O(N)$ interactions \cite{codello13, codello14,shunsuke17} or Potts model $S_{n+1}$-symmetries \cite{riccardo}. We leave all these applications to future works. 



\section*{Aknowledgments}
We would like to thank M. Safari for his help and collaboration in the early stages of this work. N.D. acknowledges support by the DFG Collaborative Research Centre ``SFB 1225 (ISOQUANT)''

\newpage
\appendix

\section{Q-functionals}

The Q--functionals appearing in the beta functionals $\beta_V$ and $\beta_Z$ of section \ref{secDE} can be evaluated analytically when the linear, or Litim, cutoff $$R_k(x) = Z_k (k^2-x) \theta (k^2-x)$$ is employed.
The explicit expressions are 
\begin{eqnarray}
Q_{n}[G_{k}^{m}\partial_{t}R_{k}] 
&=& \frac{k^{2(n-m+1)}Z_k^{1-m}}{\Gamma(n)} \left\{ (2-\eta) q_{n,m}(\zeta,\omega)+\eta q_{n+1,m}(\zeta,\omega) \right\}\nonumber\\
%
%
Q_{n}[G_{k}^{m}G'_{k}\partial_{t}R_{k}] &=& -\frac{k^{2(n-m-1)}Z_k^{-m}}{\Gamma(n)} (\zeta-1) \left\{(2-\eta) q_{n,m+2}(\zeta,\omega)+\eta q_{n+1,m+2}(\zeta,\omega) \right\}\nonumber\\
%
%
Q_{n}[G_{k}^{m}G_{k}''\partial_{t}R_{k}] &=& 2 \frac{k^{2(n-m-2)}Z_k^{-m}}{\Gamma(n)} (\zeta-1)^2 \left\{(2-\eta) q_{n,m+3}(\zeta,\omega)+\eta q_{n+1,m+3}(\zeta,\omega) \right\}\nonumber\\
&& - \frac{k^{2(n-m-2)}Z_k^{-m}}{\Gamma(n)} \frac{1}{(\zeta+\omega)^{m+2}}\,,
\end{eqnarray}
where we defined the threshold integral
\begin{equation}
q_{n,m}(\zeta,\omega) \equiv 
 \frac{1}{n(1+\omega)^m}\,{}_2F_1\!\left(m,n;n+1;\frac{1-\zeta}{1+\omega}\right)\,,
\end{equation}
with $\omega = \frac{V''_k(\phi)}{k^2 Z_k}$ and $\zeta = \frac{Z_k(\phi)}{Z_k}$.
\newpage
\section{Beta functionals}
Here we report the explicit linear cutoff expressions, in two and three dimensions, for the dimensionless beta functionals $\beta_v$ and $\beta_\zeta$ defined in section \ref{secDE}.

\subsection{$d=2$}
\begin{align}
 \beta_v =& 
 - 2 v +  \frac{\eta}{2} \, \varphi \,v'  +
  \frac{1}{8\pi} \left\{ \frac{\eta }{ \zeta -1}-\frac{2+\eta  \omega -  (2-\eta ) \zeta}{ (\zeta -1)^2}\log
\frac{\zeta+\omega}{1+ \omega} \right\} 
\label{flow_V_d=2_litim}
\end{align}
\begin{align}
 \beta_\zeta =& \, \eta \,\zeta + \frac{\eta}{2} \, \varphi \,\zeta' \nonumber\\
 +&\, \frac{ \zeta''}{8\pi} \left\{ \frac{2+\eta  \omega -  (2-\eta)\zeta}{  (\zeta -1) (1+ \omega) (\zeta+\omega )} 
  -\frac{\eta  }{ (\zeta -1)^2} \log \frac{\zeta+\omega}{1+ \omega}\right\} \nonumber\\
+&  \,  \frac{(v''')^2 }{8\pi} \left\{   \frac{2-\eta }{3(\zeta +\omega )^3}-\frac{2-\eta }{3(1+\omega )^3} -\frac{2\zeta}{(\zeta +\omega )^4}\right\} \nonumber\\
+ &   \frac{v''' \zeta' }{ 4\pi}\,\left\{ \frac{4-\eta}{(1+\omega)^2 (\zeta +\omega )^4} \omega^3 
+\frac{ \zeta  (18-7 \eta )-2\eta +12}{3(1+\omega)^2 (\zeta +\omega )^4}\omega^2  
\right.\nonumber\\
&\qquad\quad\quad+ \frac{ 2\zeta ^2 (8-3 \eta )+2\zeta  (2- \eta )+4-\eta}{3(1+\omega)^2 (\zeta +\omega )^4}\omega
\left. + \frac{2 \zeta ^3 (2-\eta)+4 \zeta ^2-\zeta  (2+\eta)}{3(1+\omega)^2 (\zeta +\omega )^4} \right\}\nonumber\\
+&  \, \frac{(\zeta')^2}{8\pi}\left\{ -\frac{3 \eta}{(\zeta -1)^2 (1+\omega) (\zeta +\omega )^4} \omega^4 - \frac{3 \eta(7 \zeta   +1 )}{2 (\zeta -1)^2 (1+\omega) (\zeta +\omega )^4} \omega^3 \right.\nonumber\\
& \qquad\quad\quad+ \frac{\zeta ^2 (30-83 \eta )-\zeta  (23 \eta +60)-2 \eta +30}{6 (\zeta -1)^2 (1+\omega) (\zeta +\omega )^4} \omega^2 \nonumber\\
& \qquad\quad\quad+ \frac{\zeta ^3 (36-47 \eta )-\zeta ^2 (25 \eta +60)+2 \zeta  (6+\eta)+12-2 \eta }{6(\zeta -1)^2 (1+\omega) (\zeta +\omega )^4} \omega \nonumber\\
& \qquad\quad\quad+ \frac{ 9 \zeta ^4 (2- \eta )-\zeta ^3 (11 \eta +36)+\zeta ^2 (18+4 \eta)
   -2 \zeta\eta }{6 (\zeta -1)^2 (1+\omega) (\zeta +\omega )^4}  \nonumber\\
 & \qquad\quad\quad \left.  +\frac{3 \eta  }{ (\zeta-1)^3} \log \frac{\zeta +\omega }{1+\omega} \right\}\,.
 \label{flow_Z_d=2_litim}
\end{align}
\subsection{$d=3$}
\begin{align} \label{flow_V_d=3_litim}
\beta_v=&  - 3 v +  \frac{1+\eta}{2} \varphi \,v'  +\frac{1}{4\pi^2}\left\{-\frac{6+(1+3  \omega)\eta-2   (3-\eta) \zeta}{3 (\zeta-1)^2}\right.
\nonumber\\&\left.
\qquad\quad+\frac{2+\eta  \omega - (2-\eta)\zeta  }{(1-\zeta )^{5/2}} \sqrt{1+\omega} \,{\rm arctanh} \sqrt{\frac{1-\zeta }{1+\omega}} \right\} 
\end{align}
\begin{align}
\beta_\zeta  = & \, \eta \,\zeta + \frac{1+\eta}{2} \,\varphi \,\zeta' \nonumber\\
+&\,\frac{\zeta''}{8\pi^2} \left\{ \frac{\zeta(2-3\eta)-2-3\eta\omega}{(\zeta-1)^{2}(\zeta+\omega)}+\frac{2+\eta(2+3\omega)-\zeta(2-\eta)}{(1-\zeta)^{5/2}\sqrt{\omega+1}}{\rm arctanh}\sqrt{\frac{1-\zeta}{1+\omega}}\right\} \nonumber\\
  + & \frac{(v''')^{2}}{(4\pi)^2}\left\{ \frac{-3\zeta(\eta-2)+\eta(\omega+4)-6}{24(1-\zeta)^{3/2}(\omega+1)^{5/2}}\, {\rm arctanh}\sqrt{\frac{1-\zeta}{1+\omega}}\right.\nonumber
 \\
   & \qquad\quad\;\;-\frac{1}{12}\frac{\eta-1}{(\zeta-1)^{2}(\omega+1)}+\frac{1}{12}\frac{\eta-1}{(\zeta-1)^{2}(\zeta+\omega)}+\frac{1}{18}\frac{\eta-2}{(\zeta-1)(\omega+1)^{2}} \nonumber\\
  & \qquad\quad\;\;\left. -\frac{2}{9}\frac{(\eta-3)}{(\zeta+\omega)^{3}}-\frac{4}{3}\frac{\zeta}{(\zeta+\omega)^{4}}+\frac{1}{12}\frac{1}{(\zeta-1)(\zeta+\omega)^{2}} \right\} \nonumber\\
  + & \frac{v'''\zeta'}{8\pi^2}\left\{ -\frac{7}{12}\frac{\zeta(\eta-2)-\eta(3\omega+4)+2}{(1-\zeta)^{5/2}(\omega+1)^{3/2}}\,{\rm arctanh}\sqrt{\frac{1-\zeta}{1+\omega}}\right.\nonumber\\
   & \qquad\quad\;\; -\frac{7}{12}\frac{\eta-2}{(\zeta-1)^{2}(\omega+1)}-\frac{4}{9}\frac{\eta-5}{(\zeta+\omega)^{3}}-\frac{8}{3}\frac{\zeta}{(\zeta+\omega)^{4}} \nonumber\\
   & \qquad\quad\;\; \left. - \frac{7}{6}\frac{\eta+1}{(\zeta-1)^{2}(\zeta+\omega)}-\frac{7}{3}\frac{1}{(\zeta-1)(\zeta+\omega)^{2}} \right\} \nonumber\\
  + & \frac{(\zeta')^{2}}{8\pi^2}\left\{ \frac{49}{24}\frac{\zeta(\eta-2)+\eta(5\omega+4)+2}{(1-\zeta)^{7/2}\sqrt{\omega+1}}\,{\rm arctanh}\sqrt{\frac{1-\zeta}{1+\omega}}\right.\nonumber\\
   & \qquad\quad\;\;-\frac{2}{9}\frac{\eta-7}{(\zeta+\omega)^{3}}+\frac{245\eta}{24(\zeta-1)^{3}}-\frac{4}{3}\frac{\zeta}{(\zeta+\omega)^{4}}\nonumber\\
   & \qquad\quad\;\; \left.-\frac{49}{32}\frac{\eta+3}{(\zeta-1)^{2}(\zeta+\omega)}-\frac{49}{18}\frac{1}{(\zeta-1)(\zeta+\omega)^{2}}\right\} \,.
 \label{flow_Z_d=3_litim}
\end{align}


\begin{thebibliography}{99}
%
\bibitem{Wilson75}
K.~G.~Wilson,
Rev.\ Mod.\ Phys.\  {\bf 47} (1975) 773;
K.~G.~Wilson,
Rev.\ Mod.\ Phys.\  {\bf 55} (1983) 583.

\bibitem{Wetterich93}
N.~Tetradis and C.~Wetterich,
Nucl.\ Phys.\ B {\bf 422} (1994) 541
[hep-ph/9308214].

\bibitem{Morris94}
T.~R.~Morris,
Phys.\ Lett.\ B {\bf 329} (1994) 241
[hep-ph/9403340];
T.~R.~Morris,
Nucl.\ Phys.\ Proc.\ Suppl.\  {\bf 42} (1995) 811
[hep-lat/9411053].

\bibitem{Berges2002}
J.~Berges, N.~Tetradis and C.~Wetterich,
Phys.\ Rept.\  {\bf 363} (2002) 223
[hep-ph/0005122].

\bibitem{Schwenk2012}
A. Schwenk, and J. Polonyi,
Lectures Notes in Physics \textbf{852} (2012).

\bibitem{Kopietz2010}
P. Kopietz, L. Bartosch, and F. Schütz, 
Lecture Notes in Physics \textbf{798} (2010).

\bibitem{Morris1996}
T.~R.~Morris,
Phys.\ Rev.\ Lett.\  {\bf 77} (1996) 1658
[hep-th/9601128].

\bibitem{codello13}
A.~Codello and G.~D'Odorico,
Phys.\ Rev.\ Lett.\  {\bf 110} (2013) 141601
[arXiv:1210.4037 [hep-th]].

\bibitem{codello14}
A.~Codello, N.~Defenu and G.~D'Odorico,
Phys.\ Rev.\ D {\bf 91} (2015) no.10,  105003
[arXiv:1410.3308 [hep-th]].

\bibitem{codello12}
A.~Codello,
J.\ Phys.\ A {\bf 45} (2012) 465006
[arXiv:1204.3877 [hep-th]].

\bibitem{morris95}
T.~R.~Morris,
Phys.\ Lett.\ B {\bf 345} (1995) 139
[hep-th/9410141].
 
\bibitem{pol}
R.~Neves, Y.~Kubyshin and R.~Potting,
hep-th/9811151.

\bibitem{alfio}
A.~Bonanno and D.~Zappala,
Phys.\ Lett.\ B {\bf 504} (2001) 181
[hep-th/0010095].

\bibitem{Canet02}
L.~Canet, B.~Delamotte, D.~Mouhanna and J.~Vidal,
Phys.\ Rev.\ D {\bf 67} (2003) 065004
[hep-th/0211055].

\bibitem{LitimZappala}
D.~F.~Litim and D.~Zappala,
Phys.\ Rev.\ D {\bf 83} (2011) 085009
[arXiv:1009.1948 [hep-th]].
 
\bibitem{Litim2000}
D.~F.~Litim,
Phys.\ Lett.\ B {\bf 486} (2000) 92
[hep-th/0005245];
D.~F.~Litim,
Phys.\ Rev.\ D {\bf 64} (2001) 105007
[hep-th/0103195].

\bibitem{Defenu15}
Defenu, N., Mati, P., M\'ari\'an, I.G. et al. J. High Energ. Phys. (2015) 2015: 141. 

\bibitem{shunsuke17}
S. Yabunaka, B. Delamotte, 
arXiv:1707.04383 [cond-mat.stat-mech].
   
\bibitem{zanussowipf}
T.~Hellwig, A.~Wipf and O.~Zanusso,
Phys.\ Rev.\ D {\bf 92} (2015) no.8,  085027
[arXiv:1508.02547 [hep-th]]..

\bibitem{Mussardo10}
G.~Mussardo,
{\it Statistical field theory: an introduction to exactly solved models in statistical physics}, Oxford University Press, 2010.

\bibitem{nagy12}
S.~Nagy, J.~Krizsan and K.~Sailer,
JHEP {\bf 1207} (2012) 102
[arXiv:1203.6564 [hep-th]].

\bibitem{osborn}
J.~O'Dwyer and H.~Osborn,
Annals Phys.\  {\bf 323} (2008) 1859
[arXiv:0708.2697 [hep-th]].

\bibitem{bologna}
A.~Codello, M.~Safari, G.~P.~Vacca and O.~Zanusso,
JHEP {\bf 1704} (2017) 127
[arXiv:1703.04830 [hep-th]].
  
\bibitem{bologna2}
A.~Codello, M.~Safari, G.~P.~Vacca and O.~Zanusso,
arXiv:1705.05558 [hep-th].

\bibitem{bologna3}
A.~Codello, M.~Safari, G.~P.~Vacca and O.~Zanusso,
Phys.\ Rev.\ D {\bf 96} (2017) no.8,  081701
[arXiv:1706.06887 [hep-th]].

\bibitem{Zambelli2015}
L.~Zambelli,
arXiv:1510.09151 [hep-th].

\bibitem{Zambelli2016}
L.~Zambelli and O.~Zanusso,
Phys.\ Rev.\ D {\bf 95} (2017) no.8,  085001
[arXiv:1612.08739 [hep-th]].

\bibitem{Defenu2015}
N.~Defenu, A.~Trombettoni and A.~Codello,
Phys.\ Rev.\ E {\bf 92} (2015) no.5,  052113
[arXiv:1409.8322 [cond-mat.stat-mech]].
  
\bibitem{Defenu2016}
N.~Defenu, A.~Trombettoni and S.~Ruffo,
Phys.\ Rev.\ B {\bf 94} (2016) no.22,  224411
[arXiv:1606.07756 [cond-mat.stat-mech]].

\bibitem{Defenu2017}
N.~Defenu, A.~Trombettoni and S.~Ruffo,
Phys.\ Rev.\ B {\bf 96} (2017) no.10,  104432.

\bibitem{riccardo}
R.~B.~A.~Zinati and A.~Codello,
arXiv:1707.03410 [cond-mat.stat-mech].
  

\end{thebibliography}
\end{document}